\def\qq{$Q^{2}$}
\def\ds{D^{\ast}}
\def\d0{D^{0}}
\def\dspm{{\ds}^{\pm}}

\def\wrang{$130 < W < 280\GeV$}
\def\qqrang{\qq${}<1\,\g2$}
\def\ptran{$p_{\perp}^D>3$\,GeV}
\def\ptrang{$p_{\perp}^{\ds}>2$\,GeV}
\def\ptranglow{$p_{\perp}^{\ds}>1.5$\,GeV}
\def\ptranghi{$p_{\perp}^{\ds}>3$\,GeV}
\def\etaran{$|\eta^D|<1.5$}
\def\etarang{$|\eta^{\ds}|<1.5$}

\def\dspm{D^{*\pm}}

\def\dss{{D_s}}

\def\pt{p_{\!\perp}}

\def\GeV{\,\textrm{GeV}}
\def\MeV{\,\textrm{MeV}}
\def\g2{\GeV^2}

\def\pb{\,\textrm{pb}}
\def\ipb{\pb^{-1}}
\def\stat{\,\textrm{(stat.)}} 
\def\syst{\,\textrm{(syst.)}}
\def\ext{\,\textrm{(ext.)}}
\def\br{\,\textrm{(br.)}}

\def\qq{$Q^{2}$}
\def\xsecra{0.41\pm0.07\stat^{+0.03}_{-0.05}\syst\pm0.10\br}
\def\ETAJ{\eta^{jet}}
 
\documentclass[cmfonts]{aipproc}
\usepackage{times}
\layoutstyle{6x9}

\SetInternalRegister\hbadness{8000} 

\newcommand\doingARLO[2][]{%
  \ifx\mmref\undefined #1\else #2\fi
}
  
\newcommand{\xgo} {x_\gamma^{\rm OBS}\ }
 
\newcommand{\dtstz}{D_2^{*0}}
\newcommand{\dplus}{D^+}
\newcommand{\dstarplus}{D^{*+}}
\newcommand{\dstprplus}{D^{*'+}}

\begin{document}
 
\title 
      {Heavy Quark Production and Spectroscopy at HERA}               
 
 
\author{Uri Karshon  
\thanks{On behalf of the H1 and ZEUS Collaborations} }
 {address={Weizmann Institute of Science, Israel} 
 }           
 
 
 
\copyrightyear  {2001}

\begin{abstract}
Production of final states containing open charm ($c$) and beauty ($b$) quarks at HERA 
         is reviewed. Photoproduction (PHP) of the charm meson
                      resonances $D^*$, $D^0$ and
$D_s$, as well as $D^*$ production in the deep inelastic scattering (DIS) regime, are measured
and compared to QCD predictions. The excited charm mesons 
                         $D_1^0 (2420)$,    $D^{*0}_2(2460)$  and
                               $D^{\pm}_{s1}(2536)$ have been observed and
the rates of charm quarks hadronising to these mesons were extracted. A search for
radially excited charm mesons has been performed. PHP and DIS
                                 beauty cross sections are higher
than expected in     next-to-leading order (NLO) QCD.
 
\end{abstract}

\date{\today}

\maketitle
 
\section{Introduction}
 
 
 
The HERA e-p collider accelerates electrons (or positrons)
          and protons to energies of $E_e =27.5\GeV$
and $E_p =920\GeV$ ($820\GeV$           
             until 1997), respectively. The H1 and ZEUS experiments are 
located at two collision points along the circulating beams. The incoming $e^{\pm}$
interacts with the proton by first radiating a virtual photon. The photon is either
quasi-real with                                          
 $Q^2 < 1\g2$ and           
   $ Q^2_{median} \approx 3\cdot 10^{-4}\g2$ (PHP regime)
                                                                   or highly  virtual
($Q^2 > 1\g2$ - DIS regime).
 
The large masses of the heavy quarks (HQ) $c$ and $b$ provide a ``hard" scale needed
for the comparison of data to QCD predictions. In leading-order (LO) QCD, two types
of processes are responsible for the PHP of HQ's: Direct photon processes, where
the photon interacts as a point-like particle with a parton from the incoming proton, and
resolved photon processes, where a parton from the photon partonic structure scatters off a
parton from the proton. Heavy
quarks ($Q$)     present in the parton distributions of the photon    
              lead to LO resolved processes,
 such as $Qg\to Qg$ (where g is a gluon),                                                
          which are called heavy flavour excitation.     
In NLO calculations, only the sum of direct and resolved processes is unambiguously
defined.
 
Two different               NLO calculations               
                            are available for comparison with
 measurements of HQ  PHP               at HERA: 1) a                  
 fixed-order (``massive")    
approach~\cite{Frixione,Ellis}, where     HQ's                  
 are produced only dynamically                      
                              in the hard subprocess.
This calculation is expected to become less accurate
 when $\pt^2 > > m_Q^2$, where $\pt$ and $m_Q$ are the transverse momentum and mass                       
                              of the HQ;                      
2) a resummed (``massless")
approach~\cite{kniehl,cacciari}, where the massless HQ's from the photon and proton
parton distributions are used explicitly.
This calculation is expected to yield better
 results as                         
      $\pt^2 > > m_Q^2$.                                                  

 
\begin{figure}
 \hspace*{-0.4cm}                                                   
 \vspace*{-0.2cm}                                                   
  \resizebox{12.5pc}{!}{\includegraphics{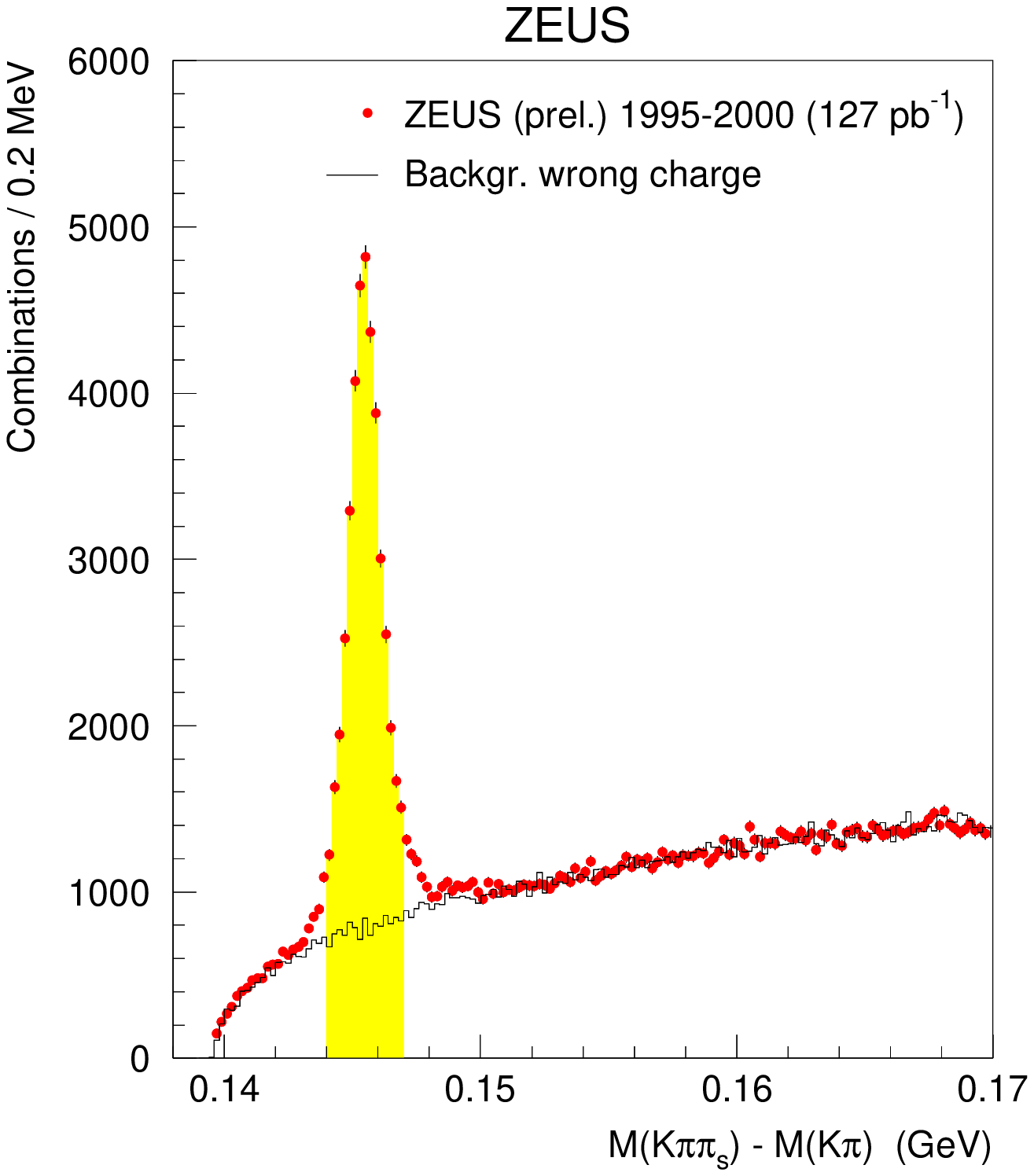}}
\vspace*{2.0cm}\hspace*{-0.9cm}\resizebox{14pc}{!}{\includegraphics{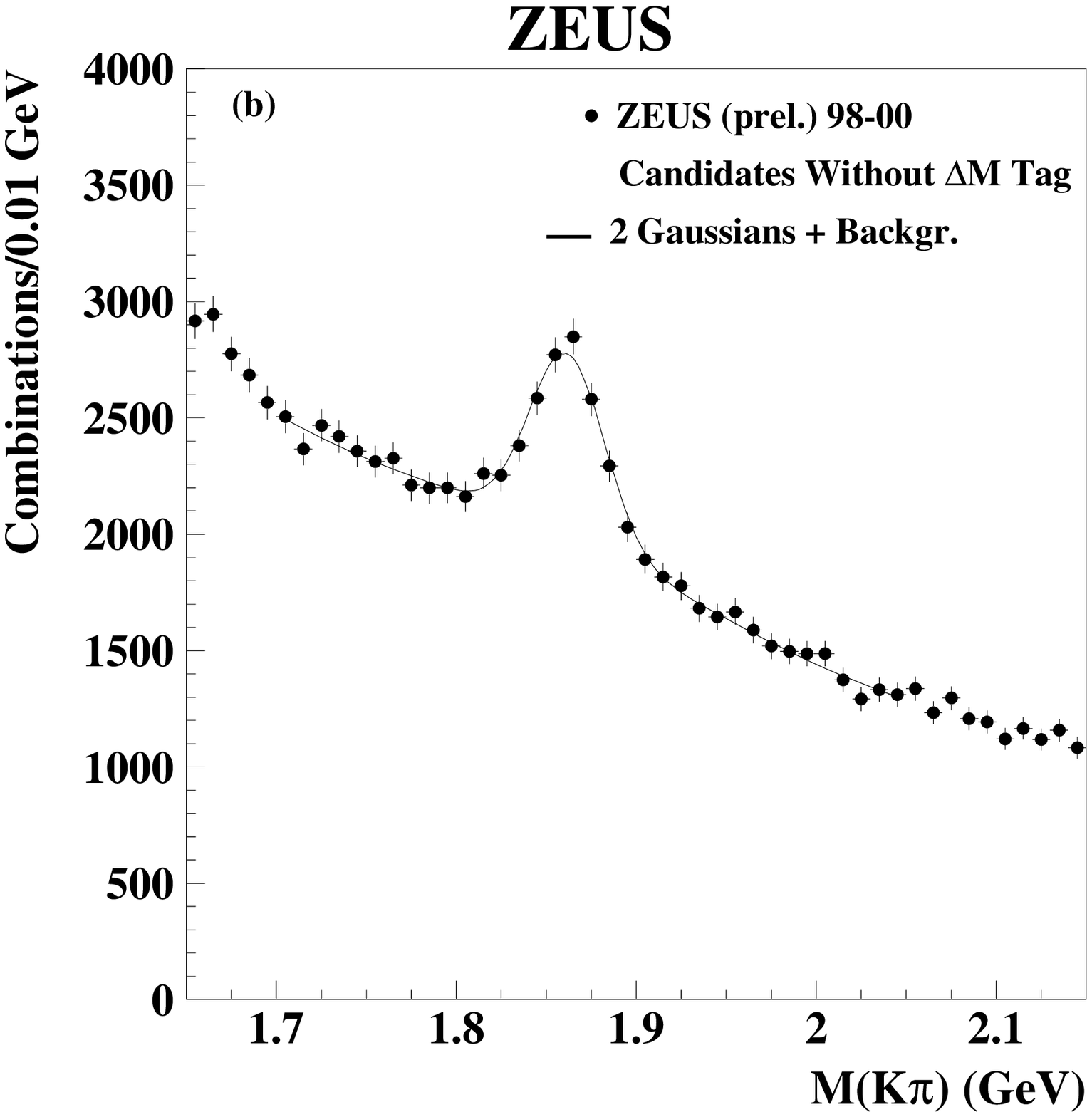}}  
\hspace*{-0.9cm}
 
\vspace*{1.8cm}
   \resizebox{12.8pc}{!}{\includegraphics{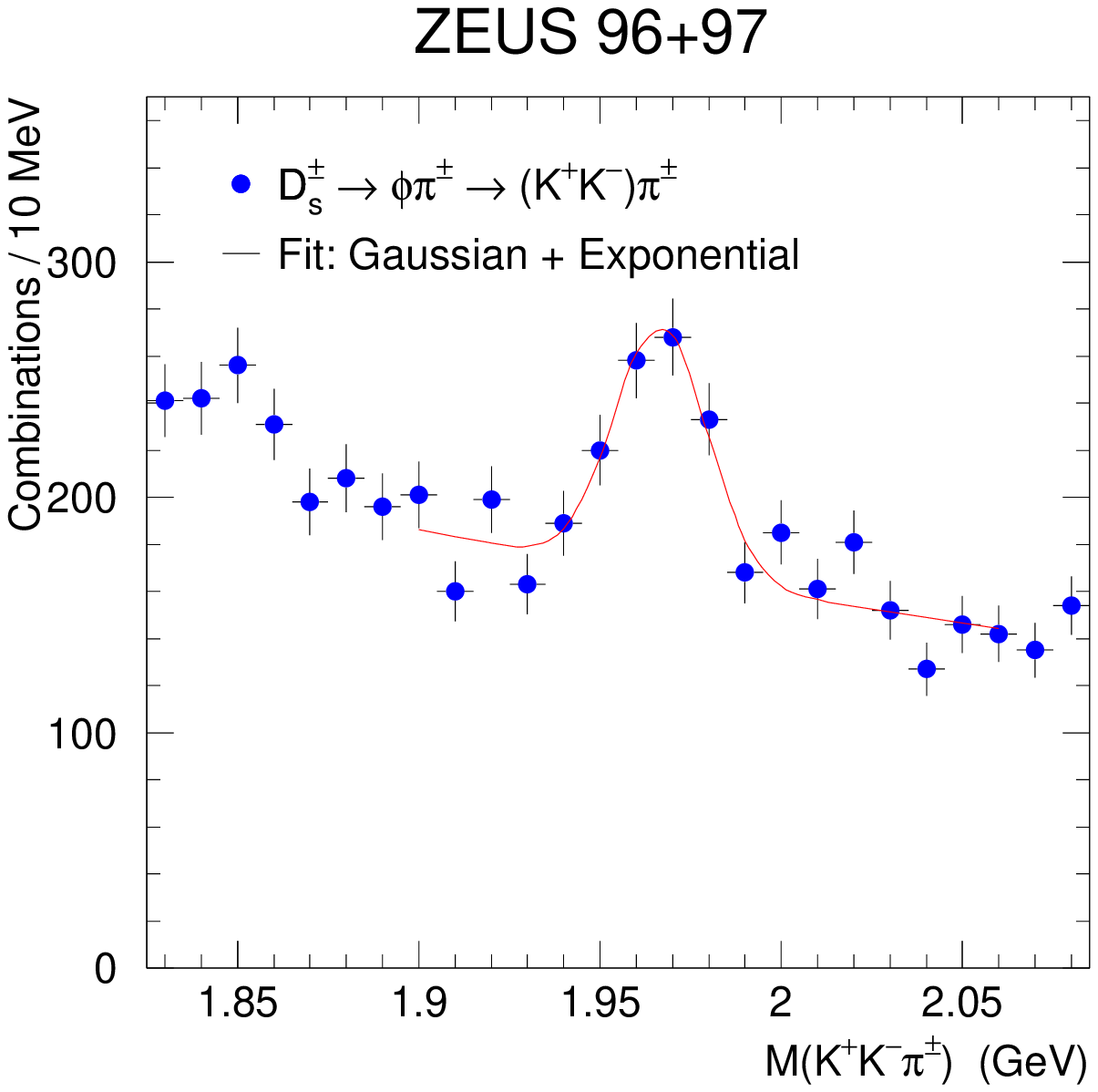}}
\caption{(a)                                             
            $M(K\pi\pi_S)-M(K\pi)$ distribution in the $D^0$ mass region (full dots).
The        histogram is the            distribution for wrong charge combinations. 
 (b) $M(K^{\pm}\pi^{\mp})$ distribution for events excluding the $\dspm$ region.
             The  solid curve is a fit
to two Gaussian shapes for the right and wrong $K$ mass assignments     
                       plus a sum of exponential and linear backgrounds.
 (c) $M(K^+ K^-\pi^{\pm})$ distribution for events inside the $\phi$ mass range
($1.0115 < M(K^+ K^- ) < 1.0275\GeV$). The solid curve is a fit to a Gaussian resonance
plus an exponential background.
        }
\end{figure}
 
\section{Production of $\dspm$, $D^0$ and $D_s$ mesons}
 
 \vspace*{-10.0cm}\hspace*{ 0.6cm}{\tiny  (a)}
 \vspace*{+10.0cm}\hspace*{-0.6cm}      
 
 \vspace*{-10.1cm}\hspace*{ 9.8cm}{\tiny   (c)}
 \vspace*{+10.1cm}\hspace*{-9.8cm}      
 
 \vspace*{-1.0cm}

The charmed meson $D^{*\pm}$ has been reconstructed                                     
                                                    via its decay chain                                          
      $D^{*+} \rightarrow D^0 \pi_S^+ \rightarrow ( K^- \pi^+ ) \pi_S^+$ (+ c.c.).  
Fig.~1(a) shows the      mass difference distribution, $\Delta M~=~M(K\pi\pi_S)-M(K\pi)$,
                     in the $D^0$ mass region $1.83 < M(K\pi) < 1.90\GeV$
                            in the kinematic range
\ptrang~and \etarang , where $p_{\perp}$ is the transverse momentum
           and  $\eta =-\ln\tan(\theta /2)$ is the pseudorapidity.   
The polar angle, $\theta$, is defined with respect to the proton beam direction.
 The plot includes PHP and DIS ZEUS data collected during            
                                   1995-2000~\cite{ds1}.
A clear $\dspm$ signal is seen (dots) on top of a small combinatorial background, estimated
by wrong charge combinations (histogram), where both $D^0$ tracks have the
same charge and $\pi_S$ has the opposite charge.                                  
Defining $\dspm$ candidates as events with $0.144 < \Delta M < 0.147\GeV$,
                            a signal of $31350\pm 240~\dspm$                         
                                                             mesons was found          
      after background subtraction.
 
                    Inclusive production of charm hadrons other than 
the $\dspm$                                                                             
have also been observed. 
The $M(K^{\pm}\pi^{\mp})$ distribution for PHP events excluding the $\dspm$ region
                                          ($0.143 < \Delta M < 0.148\GeV$) 
       is shown in Fig.~1(b) for a restricted kinematic region,
using         a ZEUS           event sample with integrated luminosity                   
${\cal L} = 66\ipb$~\cite{pv}.                      
A clear $D^0$ signal is seen.                                                            
                    The production ratio, $P_v$, of vector to pseudoscalar+vector
ground state (orbital angular momentum $L=0$ of the $c\bar q$ system)
charm mesons can be approximated by 
$P_v = (\sigma(D^0)/\sigma(D^{*+}) -  B_{D^{*+}\to D^0\pi})^{-1}$, where 
$\sigma(D^0)$ and $\sigma(D^{*+})$ are, respectively, the     
                          inclusive $D^0$ and $\dspm$ cross sections 
and $B_{D^{*+}\to D^0\pi}$ is the $D^{*+}$ branching ratio to $D^0\pi^+$.
 Using also the $D^0$ signal from events in the $D^*$ region   
 with the same cuts yields the preliminary result
 $P_v=0.546\pm 0.045(\rm{stat.})\pm 0.028(\rm{syst.})$. This measurement is in good agreement
with the $e^+ e^-$ annihilation results, $0.57\pm 0.05$ and $0.595\pm 0.045$~\cite{pvlep},
supporting the universality of charm fragmentation.
 
Using a ZEUS PHP     sample with $38\ipb$
in a restricted kinematic region, a $D_s$ signal is seen (Fig.~1(c)) in the $K^+ K^-\pi^{\pm}$
mass distribution for events where $M(K^+ K^-)$ is in the $\phi$ mass range~\cite{D_s}.
The ratio of the $D_s$ to $D^*$ cross sections in identical
kinematic regions was measured to be
\( \sigma_{ep\to\dss X}/   
 \sigma_{ep\to\ds X} =\xsecra \), where the last error is due to     
                                       the uncertainty in the     
$D_s\to\phi\pi$ branching ratio. This result is in good agreement with the ratio
\( f(c\rightarrow~D_s^{+})~/~
 f(c\rightarrow~D^{*+})~=~0.43\pm~0.04\pm~0.11\br \) obtained~\cite{D_s,lgcomp}
 from $e^+ e^-$ experiments, again confirming the universality of charm fragmentation.
 
\begin{figure}
 
 \vspace*{+1.2cm}                                                   
 \hspace*{-0.9cm}                                                   
                                                                        
 \resizebox{13.0pc}{!}{\includegraphics{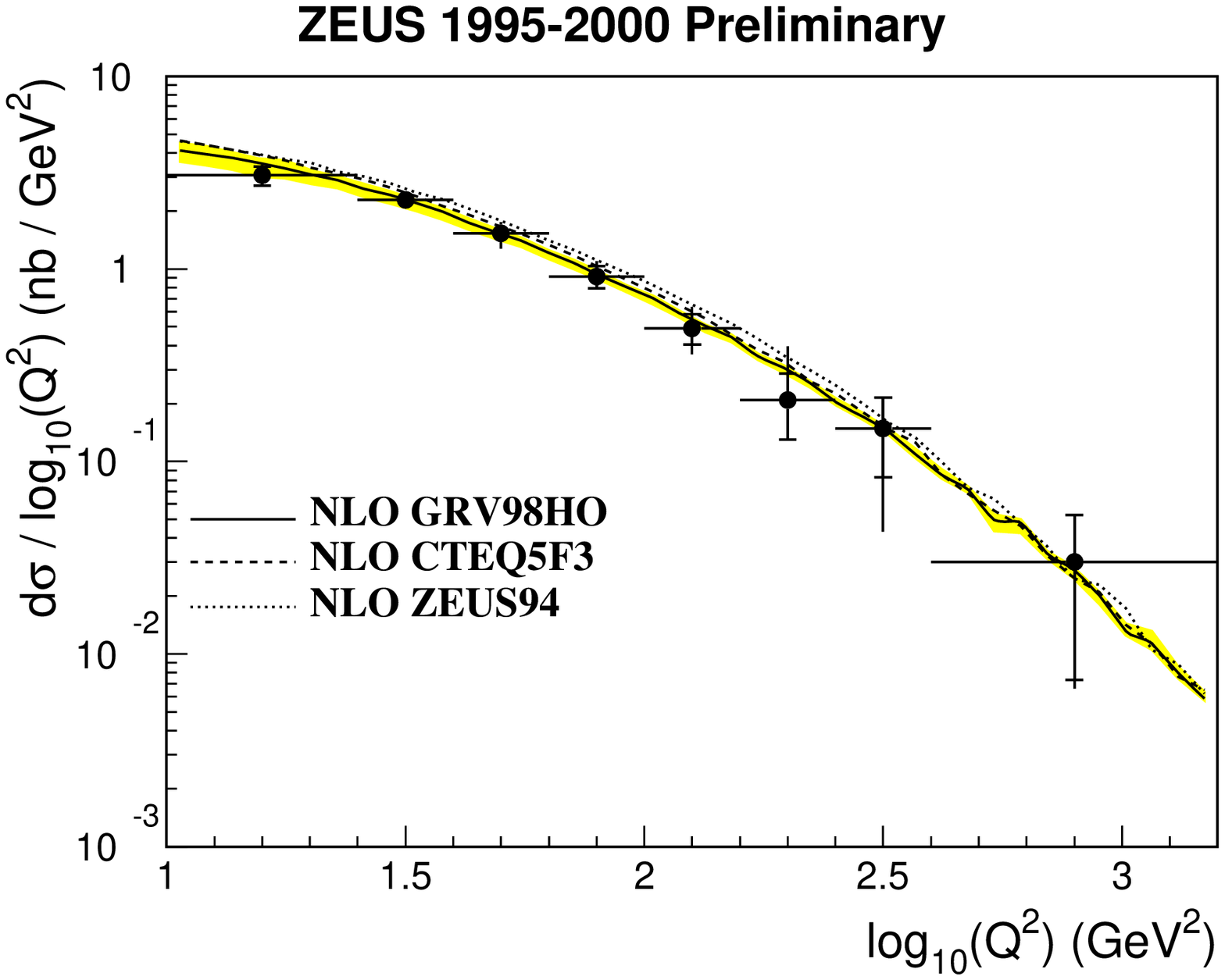}}
\vspace*{2.0cm}\hspace*{-0.1cm}      
   \resizebox{ 9.5pc}{!}{\includegraphics{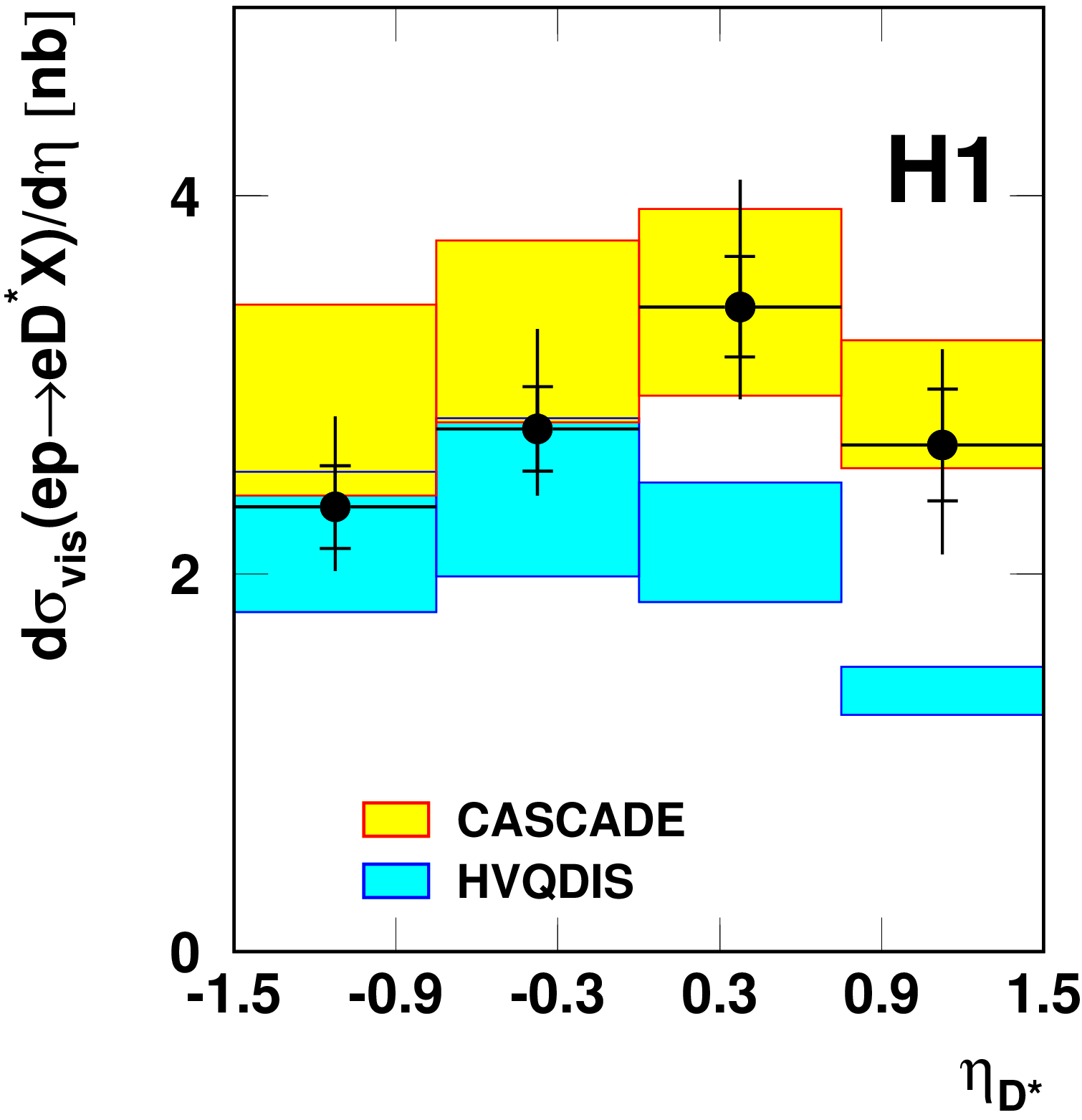}}          
 
\hspace*{ 0.0cm}
\vspace*{1.8cm}
 \resizebox{11.0pc}{!}{\includegraphics{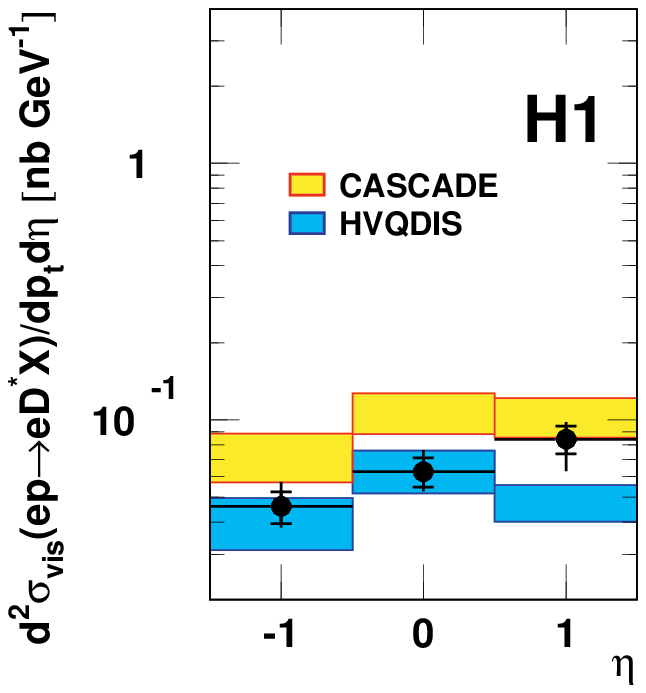}}       
 
\caption{ (a)                                            
Differential $D^*$ cross section  in $\log_{10}(Q^2)$                         
compared to the NLO HVQDIS calculations with different structure function parametrisations.
(b)        Differential $D^*$ DIS ($1 < Q^2 < 100\g2$)
cross section  in $\eta^{D^*}$ compared with HVQDIS (lower shaded band)
and CASCADE (upper shaded band) predictions.
(c) Same as b) for $4 < \pt^{D^*} < 10\GeV$.
        }
 
\end{figure}
 
 \vspace*{-10.5cm}\hspace*{ 2.2cm}{\tiny  (a)}
 \vspace*{+10.5cm}\hspace*{-2.2cm}      
 \vspace*{- 9.3cm}\hspace*{ 5.7cm}{\tiny  (b)}
 \vspace*{+ 9.3cm}\hspace*{-5.7cm}      
 \vspace*{+0.2cm}\hspace*{10.3cm}{\tiny   (c)}
 \vspace*{-0.2cm}\hspace*{-10.3cm}      
 
 \vspace*{-0.8cm}                                                   
 
\subsection{$D^{*\pm}$ Production in DIS}
 
 \vspace*{-0.1cm}                                                   
 
Open charm production in the DIS regime is dominated by boson-gluon fusion (BGF) processes,      
 where the boson (gluon) is emitted from the incoming electron (proton).
Fixed-order NLO perturbative QCD (pQCD) calculations are available in the form of a   
Monte Carlo (MC) integrator (HVQDIS)~\cite{hvqdis}.                                       
                      The ZEUS preliminary          $D^*$ differential cross section 
in $Q^2$,                     using a sample of ${\cal L}=82.6\ipb$, is  shown in 
Fig.~2(a) for the kinematic region $ Q^2 > 10\g2$,
\ptranglow~and \etarang~\cite{osa855}.                                  
                  The distribution compares well with the HVQDIS calculations,
using 3 different parton distribution functions in the proton and a charm quark mass range
$1.3 < m_c < 1.6\GeV$. For $Q^2 > > m_c^2$, resummed NLO calculations should be superior.
However, up to $Q^2\approx 1000\g2$,                    the data is nicely described by
the                  fixed-order                        scheme. 
The $\dspm$ cross sections were also measured separately                                 
              for $e^+$ and $e^-$ beams~\cite{bu493}.
Integrated over $Q^2 > 20\g2$, the $e^- p$ cross section is higher than that 
   for $e^+ p$ by 
$\approx 3$ standard deviations.
Both results are compatible with the predictions within the theoretical uncertainties.
 
The H1 Collaboration has measured $\dspm$ production in DIS, using                
${\cal L} = 18.6\ipb$, in a kinematic region $1 < Q^2 < 100\g2$,
\ptranglow~and \etarang~\cite{H1dis}. 
   The differential cross section in $\eta^{D^*}$ is compared in
Figs.~2(b-c) to HVQDIS~\cite{hvqdis} and to CASCADE~\cite{cascade} calculations, which
implement a version of the CCFM evolution scheme~\cite{CCFM}. The shaded
bands reflect the uncertainties in the predictions due to $m_c$ ($1.3-1.5\GeV$) and
the allowed fragmentation parameters. CASCADE shows better agreement with the overall
$\eta^{D^*}$ distribution (Fig.~2(b)),
while the HVQDIS prediction is too low for the forward region. CASCADE agrees poorly
with the data at the low-$\eta^{D^*}$     high-$\pt^{D^*}$ region (Fig.~2(c)).
 
\begin{figure}
 
 \vspace*{ 0.5cm}                                                   
 \hspace*{-2.0cm}                                                   
  \resizebox{14.0pc}{!}{\includegraphics{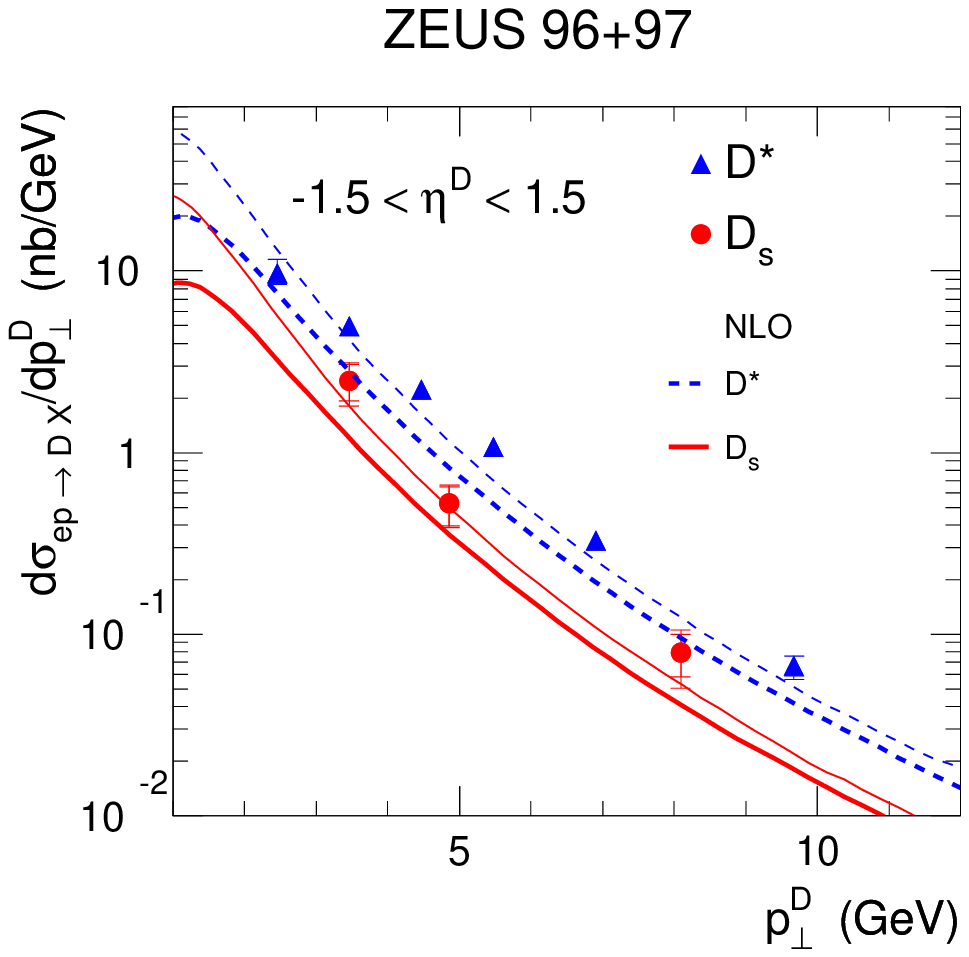}}
 
\vspace*{2.0cm}\hspace*{+0.9cm}\resizebox{14pc}{!}{\includegraphics{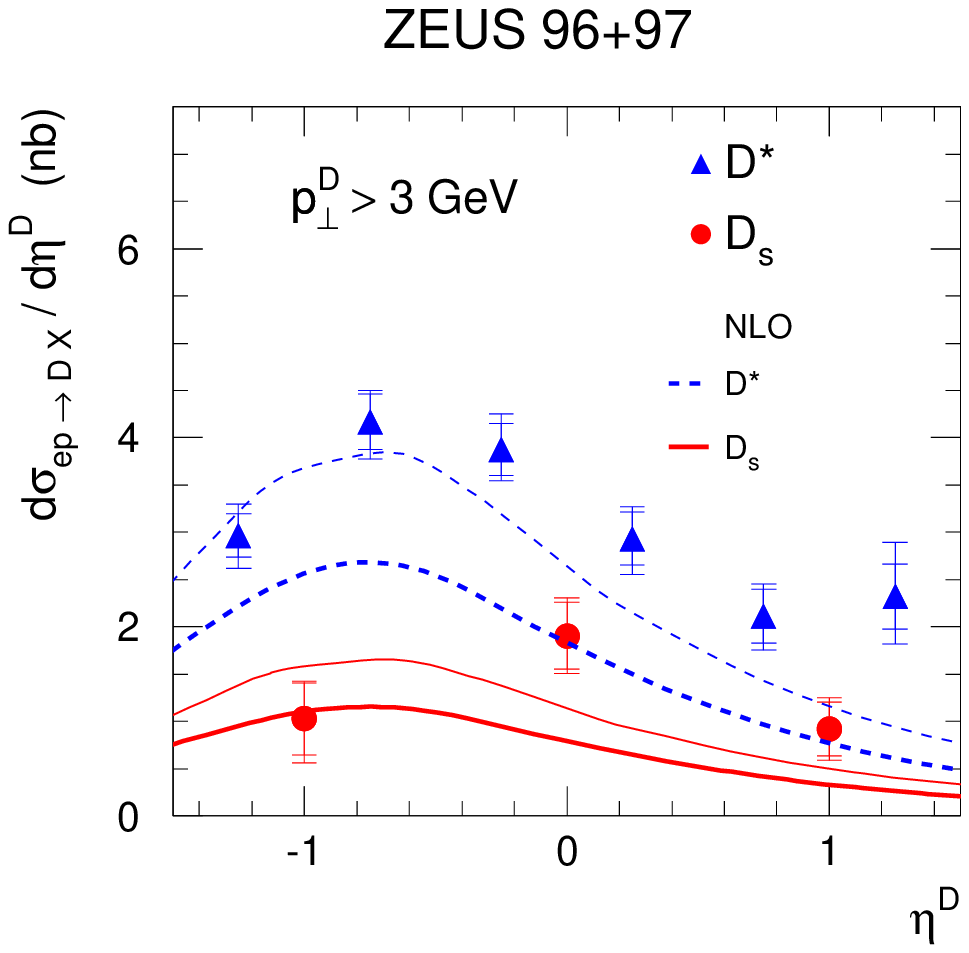}} 
\caption{ Differential cross sections in              
                                  $\pt^{D}$ and
      $\eta^{D}$, where \protect\( D\protect \) stands
  for \protect\( \ds \protect \) or \protect\( \dss \protect \).    
                 The \protect\( \dss \protect \) (dots) and \protect\(
  \ds \protect \) (triangles) data are compared with     NLO
  predictions for $\dss$ (full curves) and $\ds$ (dashed curves) with
  two parameter settings: \( m_{c}=1.5\GeV \), \( \mu _{R} = m_{\perp
    } \) (thick curves) and \( m_{c}=1.2\GeV \), \( \mu _{R} =
  0.5m_{\perp } \) (thin curves). Here $\mu_R$ is the renormalisation scale 
and $m_{\perp }=\sqrt{m_c^2 + p_{\perp}^2}$.        
        }
\end{figure}
 
 \vspace*{-0.4cm}                                                   
 
\subsection{Photoproduction of charm mesons}
 
 \vspace*{-0.1cm}                                                   
 
Differential $D_s^{\pm}$ cross sections in $\pt$ and $\eta$~\cite{D_s} were compared
to   those for $\dspm$~\cite{dstarrf} in the same kinematic region               
                        \qqrang, \wrang,
\ptran~and \etaran~, where $W$ is the $\gamma p$ centre-of-mass energy,
                                                                   and with fixed-order NLO              
calculations~\cite{frix} (Fig.~3). The cross sections for both cases        are above
the predictions,
        in particular for $\eta$ in the forward (proton) direction.
NLO resummed predictions~\cite{kniehl}                                     
 are closer to the $D^*$ data~\cite{dstarrf},
but still too low for high $\eta$.                                           
           Using different photon parton density functions shows some sensitivity 
to the parton density parametrisation of the photon. 
  A tree-level pQCD calculation~\cite{BKL}, where the $c\bar q$ state is hadronised rather
than the $c$ quark, gives a better agreement with the data compared to the NLO
calculations.
 
 \vspace*{-0.4cm}                                                   
 
\subsection{$\dspm$ and Associated Dijets}
 
 
 \vspace*{-0.2cm}                                                   
 
Events with a reconstructed $\dspm$ and at least 2 hadron jets
(``dijet event") enable one to study the photon structure,    in 
particular its charm content. $\dspm$ candidates with \ptranghi~and 
\etarang~ have been selected. The two jets with the highest
transverse energy, $E_T$, in the pseudorapidity range 
              $|\ETAJ| < 2.4$ are required to have 
  $E_T^{jet1}$ and
  $E_T^{jet2}$ above certain values.
The fraction of the photon momentum participating in the dijet
production, $\xgo$, is defined as
$x_\gamma^{\rm OBS} = \frac{\Sigma_{\rm jets} E_T e^{-\eta}}{2yE_e}$,
where $y$ is approximately the fraction of     incoming electron
energy carried by the photon. In LO QCD, direct processes have
$\xgo=1$ while resolved processes have $\xgo < 1$. 
Samples enriched with direct (resolved) events are separated by a cut $\xgo>0.75(<0.75)$.
 
Differential cross sections in $\xgo$ for ZEUS measurements
with ${\cal L}=37\ipb$ are shown in Fig.~4(a-b)~\cite{dstarrf} and compared with LO MC simulation 
and NLO fixed-order calculation. The peak at high $\xgo$ is due to the LO-direct BGF   
process. The low $\xgo$ tail comes from LO-resolved processes, dominated by photon
charm excitation. The shape of the $\xgo$ distribution is in good agreement with the
MC simulation with $\approx 40\%$ resolved contribution (Fig.~4(a)). The fixed-order NLO calculation
lies below the data at low $\xgo$ values (Fig.~4(b)). 
                              This could be due to the fact that
no explicit charm excitation component exists in this calculation.
 
\begin{figure}
 
 \vspace*{-1.0cm}                                                   
 \hspace*{-0.3cm}                                                   
 \hbox{
  \hbox{\resizebox{19.0pc}{!}{\includegraphics{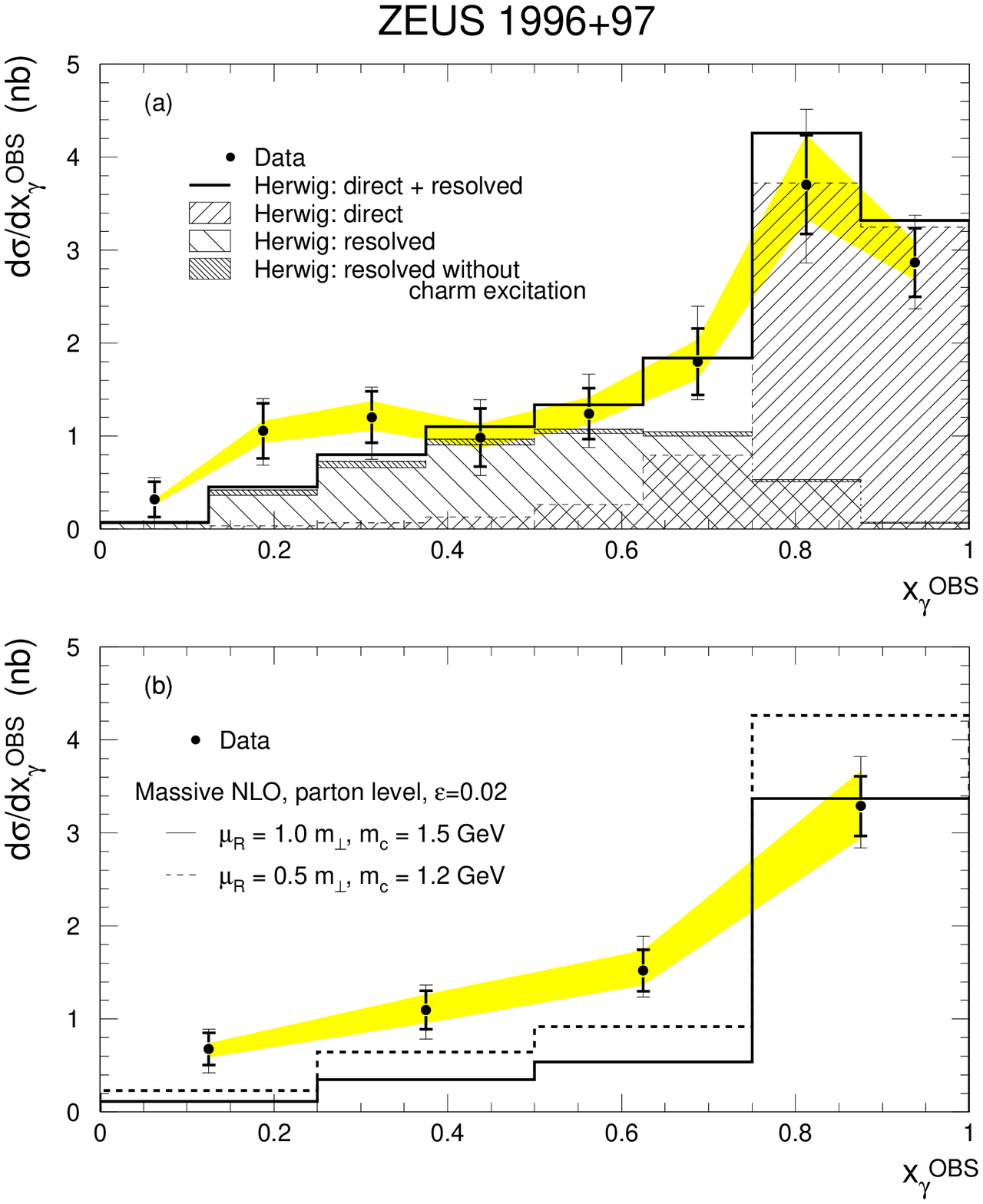}}}    
 \vbox{
 \hbox{\resizebox{16.5pc}{!}{\includegraphics{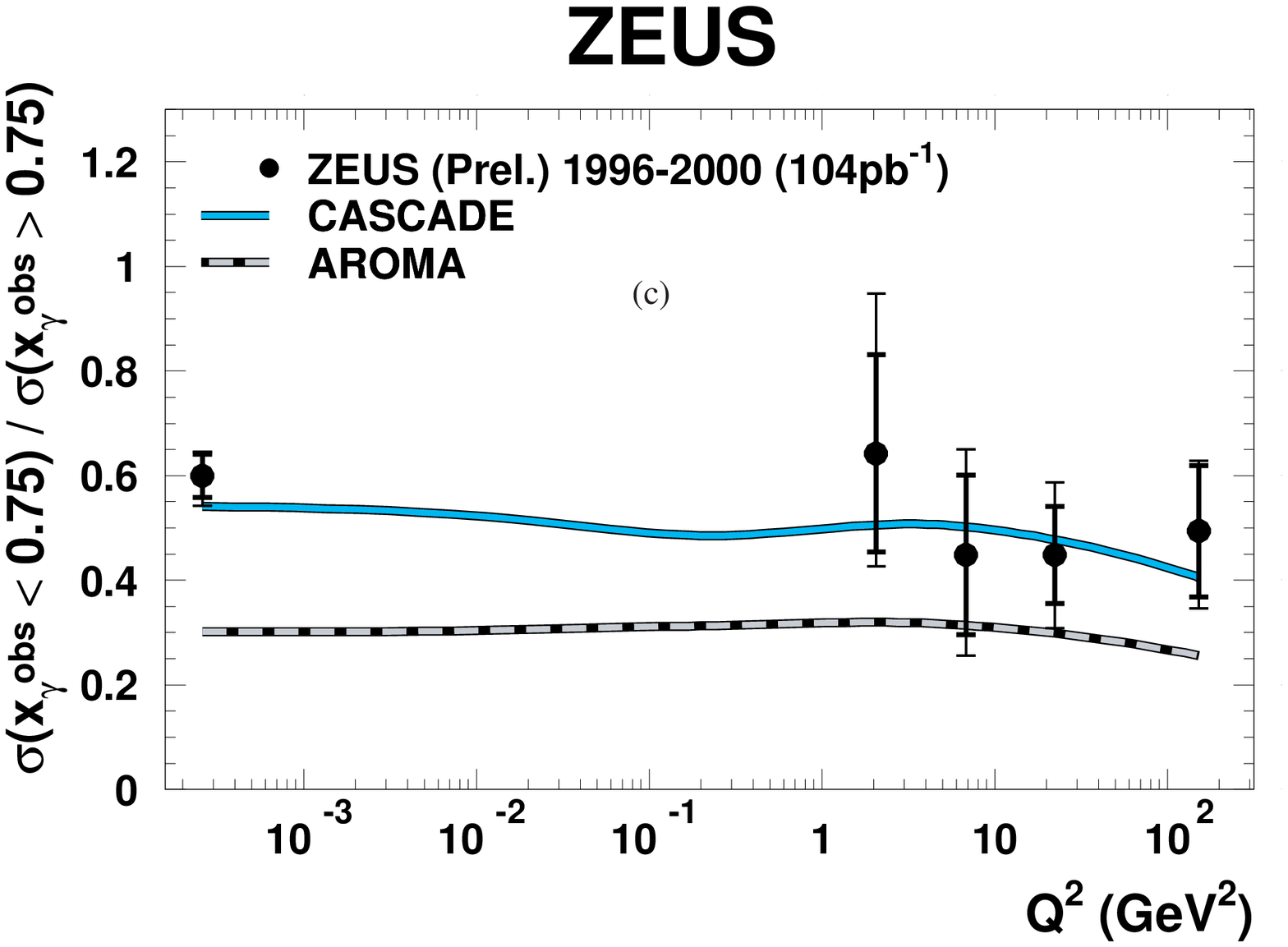}}}     
  \hbox{\resizebox{15pc}{!}{\includegraphics{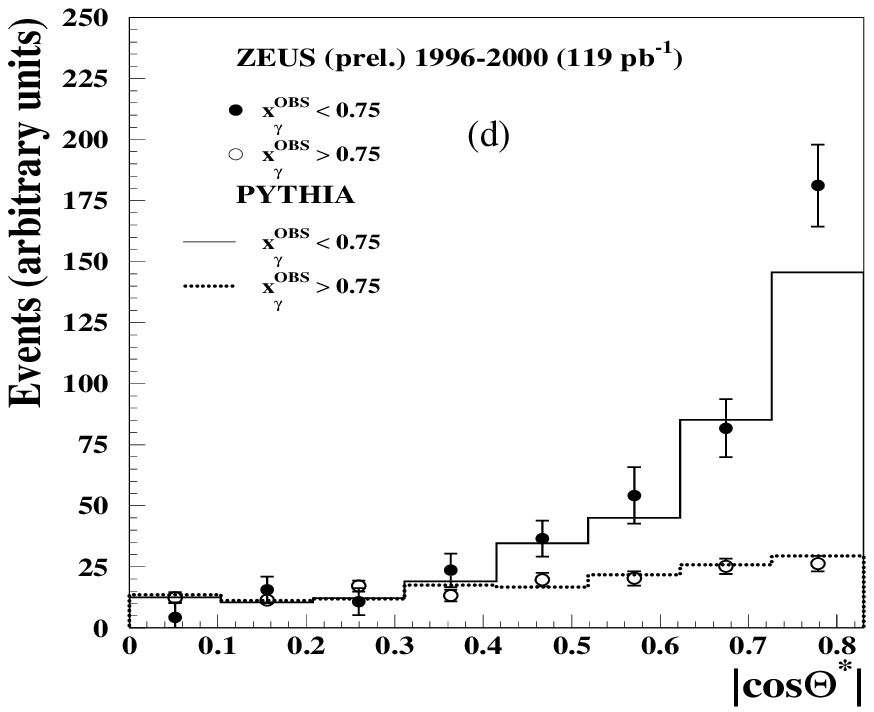}}}
}}

\caption{ (a-b) Differential cross sections    $d\sigma/d\xgo$ for dijets
  with an associated $D^*$      
                      in the PHP       range with          
   $E_T^{jet1} > 7\GeV$ and $E_T^{jet2} >
  6\GeV$.                                                              
                 In (a), the experimental data (dots) are compared to
  the expectations of the HERWIG MC simulation, normalised to the data,
  for LO-direct (right hatched), LO-resolved (left hatched),
  LO-resolved without charm excitation (dense hatched) and the sum of
  LO-direct and LO-resolved photon contributions (full histogram).  In
  (b), the data are compared with a parton level NLO fixed-order  
  calculation.                                                    
  (c) Ratio of low to high $\xgo$~for
          $D^*$ dijet events with
   $E_T^{jet1,2} > 7.5,~6.5\GeV$              
                     vs. $Q^2$                                         
                               compared to the AROMA and
  CASCADE MC's. The left-hand point is due to PHP events.
  (d) Differential distributions in $|\cos\theta^*|$
      for $D^*$ dijet events with
   $E_T^{jet1,2} > 5\GeV$.               
                    Data (dots) and PYTHIA MC            (lines)
are shown separately
  for direct- (open dots/dashed lines) and resolved- (black dots/full lines)
photon
  events. All distributions are normalised to the resolved data     in the lowest 4 bins.
        }
 
\end{figure}
 
 
The dependence of the virtual-photon structure on its virtuality, $Q^2$, has been studied
with ZEUS dijet events containing a $D^*$~\cite{ben}.
The cross section ratio $R=\sigma (\xgo < 0.75)/
                           \sigma (\xgo > 0.75)$ as a function of $Q^2$ is shown in
Fig.~4(c) to be   approximately constant up to $Q^2\approx 200\g2$,                
                                                                       contrary to the
no-charm-tag case, where $R$ falls with increasing $Q^2$.                        
                         The data are compared to two MC models
which implement no specific partonic structure for the photon, generating all low $\xgo$
events from parton showers in two different schemes. The AROMA~\cite{aroma} model, which
implements the DGLAP evolution scheme~\cite{DGLAP}, lies below the data. The CASCADE
results~\cite{cascade}                                                                
       are much closer to the data.
 
The angular distributions,                                                             
 $dN/d|\cos \theta^* |$,                              
                          of      dijet events containing a $D^*$ in the PHP regime
                         have been measured by ZEUS~\cite{cost}.                  
   Here 
 $\theta^*$ is the angle between the jet-jet axis and the beam direction in the dijet rest frame.
The results are shown in Fig.~4(d) separately for direct and resolved photon events.       
                                             The resolved processes peak                at high
 $|\cos \theta^* |$, in agreement with LO     MC predictions, as expected for dominant gluon
exchange.          The direct processes are much         flatter in
 $|\cos \theta^* |$, consistent with quark exchange. The steep rise towards high             
 $|\cos \theta^* |$        of the resolved       charm events provides     evidence that  
the bulk of the resolved contribution is due to charm excitation in the photon.

 \vspace{-0.3cm}
\section{Excited Charm Mesons}
 
P-wave charm mesons     (L=1 of the $c\bar q$       
system)      
can decay into L=0 states plus      a $\pi$ or a $K$~\cite{PDG}.
They are predicted~\cite{HQET} to                      
       appear      in two doublets with total angular momentum       
j=3/2 (narrow states) or                             
    j=1/2 (broad  states).                               
Narrow states,                           $D_1(2420)$ and $D^*_2(2460)$,
were observed in the $D^*\pi$ decay mode                  
              and identified as members of the j=3/2 doublet with
spin-parity $J^P =1^+$ and $2^+$, respectively~\cite{PDG}.
A charm-strange excited meson, $D^{\pm}_{s1}(2536)$, was found in the
$D^{*\pm}K^0_S$ final state~\cite{PDG}.     
 
\begin{figure}[t]
 \vspace{-0.8cm}
 \hspace{-0.7cm}
  \resizebox{17pc}{!}{\includegraphics{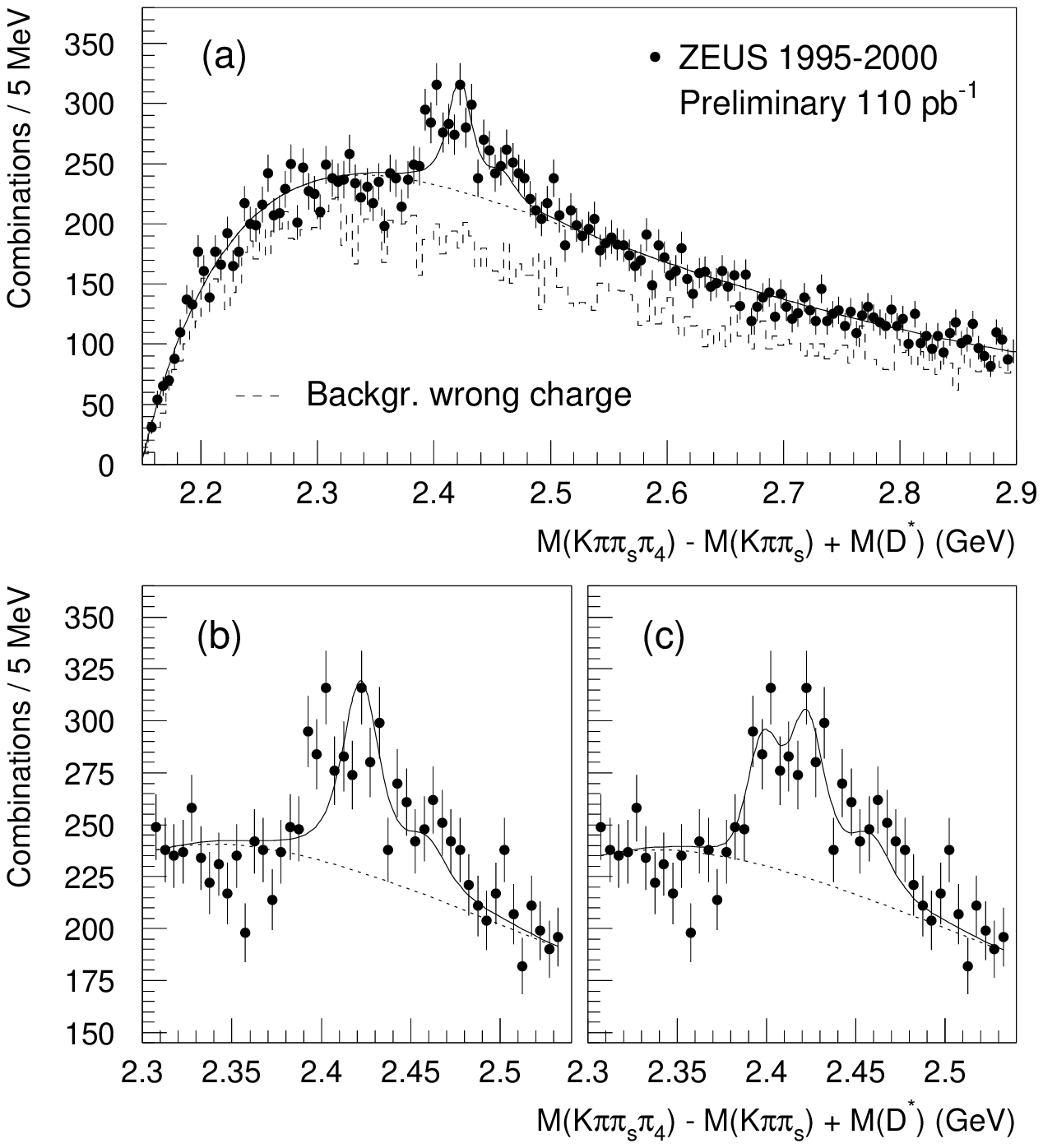}}      
 \vspace{+4.8cm}
 \hspace*{ 0.0cm}                                                   
  \resizebox{16pc}{!}{\includegraphics{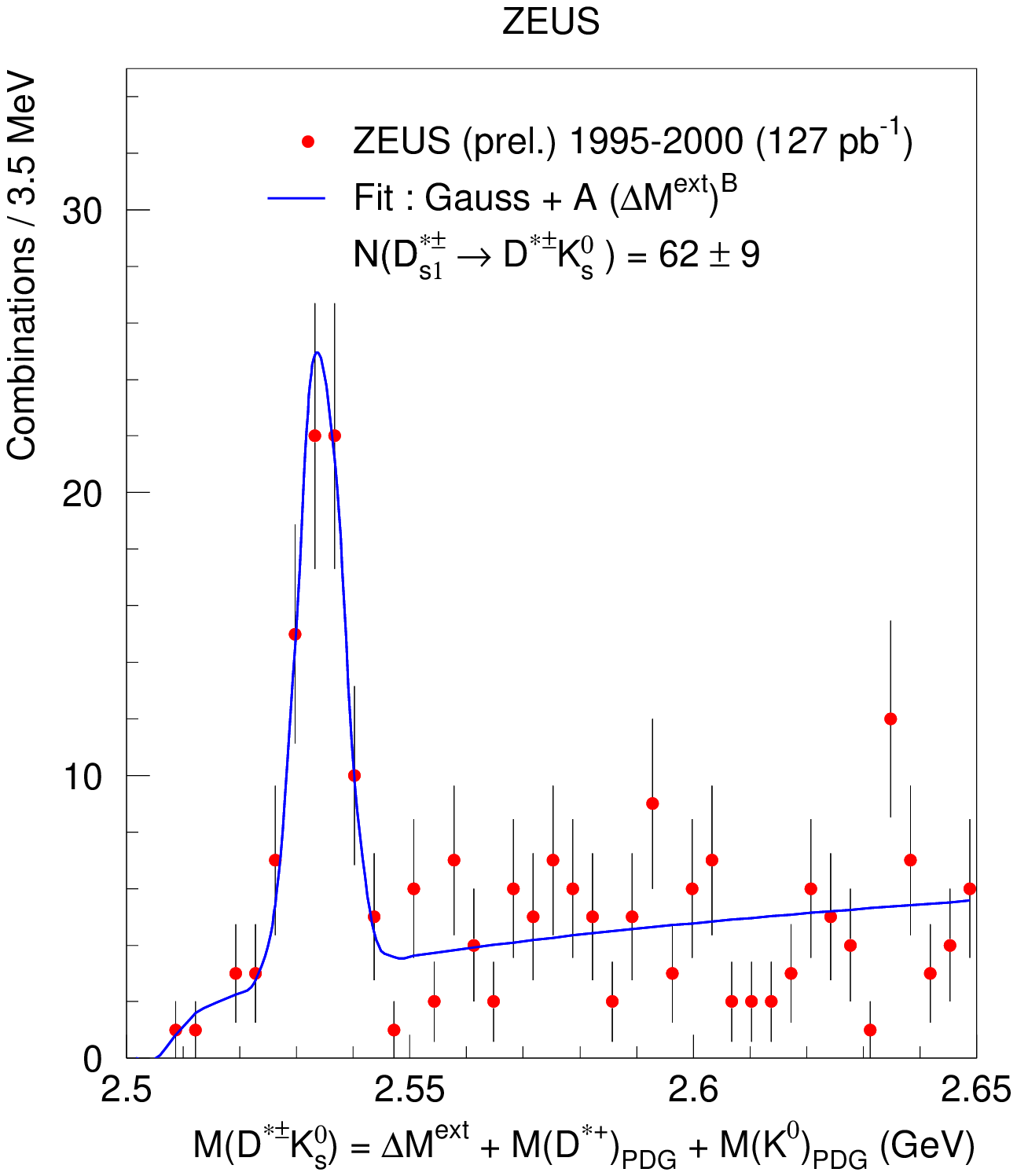}}      
 \hspace*{-1.0cm}                                                   
 \vspace{-5.5cm}
\caption{(a-c)~Extended mass difference distribution for $\dspm$ candidates (full dots).
The dashed histogram is for wrong charge cominations. The curves are the results of unbinned   
likelihood fits. In (a) and (b) the solid curves are a fit to background parametrisation
and two Breit-Wigner distributions convoluted with a Gaussian function. In (c)
an additional Gaussian-shaped resonance near $2.4\GeV$ is assumed in the fit. The dotted
curves are fitted shapes of the combinatorial background.
(d)      Effective                                                                 
          $M(\dspm K^0_S)$ distribution (full dots). The solid line is a                  
fit to a Gaussian resonance plus background of the form $A(\Delta M^{ext})^B$.        
         }
\end{figure}
 
 \vspace*{-0.5cm}                                                   
 
\subsection{$D^0_1(2420)$ and $D^{*0}_2(2460)$ production}
 
 \vspace*{-0.2cm}                                                   
 
$D^0_1$ and $D^{*0}_2$ mesons were reconstructed by  
                        ZEUS~\cite{pwave} via their decays 
to $D^{*\pm}\pi^{\mp}_4$, followed by the $D^{*\pm}$ decays,
   $D^{*+}\to D^0\pi^+_S\to (K^-\pi^+ )\pi^+_S(+c.c.)$.             
                        Fig.~5(a) shows the ``extended" mass difference distribution,
$M(K\pi\pi_S\pi_4)-M(K\pi\pi_S)+M(\ds)$, where $M(\ds)$ is the nominal $\dspm$
mass~\cite{PDG} (full dots). A clear excess is seen around the mass region of the
 $D_1^0(2420)$ and $D^{*0}_2(2460)$ mesons.                    
No enhancement is seen for wrong charge combinations (dashed histogram), where    
$D^*$ and $\pi_4$ have the same charges.
The solid curves in Figs.~5(a-b) are an unbinned likelihood fit to two
Breit-Wigner shapes with masses and widths fixed to the nominal 
$D^0_1$ and $D^{*0}_2$ values~\cite{PDG}, convoluted with a Gaussian function and
multiplied by helicity spectrum functions for $J^P=1^+$ and $2^+$ states, respectively.
                              The background shape was parametrised by the form
 $x^\alpha \cdot exp(-\beta \cdot x + \gamma \cdot x^2)$, where   
  $x =                                          
 M(K\pi\pi_S\pi_4)-M(K\pi\pi_S)-M(\pi)$. The fitted curves describe the distribution 
reasonably well, except for a narrow enhancement near
$2.4\GeV$ (Fig.~5(b)).
In Fig.~5(c), a similar fit is shown with an additional Gaussian-shaped resonance
with free mass and width.                                                             
                                                  The fit yielded $211\pm 49$ entries
for the narrow enhancement with mass value $2398.1\pm 2.1\stat ^{+1.6}_{-0.8}\syst $
\MeV. The width was consistent with the          resolution expected from the tracking
detector.                   The enhancement may
indicate a new excited charm meson, a result of an interference effect or a statistical
fluctuation.
The number of reconstructed 
$D^0_1$ and $D^{*0}_2$ mesons in the 3-resonance fit are $526\pm 65$ and $203\pm 60$,
respectively.
 
 \vspace*{-11.2cm}\hspace*{ 8.5cm}{\tiny   (d)}
 \vspace*{+11.2cm}\hspace*{-8.5cm}                      
 
The acceptance-corrected fractions of $\dspm$ mesons originating from                                       
$D^0_1$ and $D^{*0}_2$                                                in the measured
kinematic range were found to be 
 $R_{D_1^{0}\to D^{* \pm}\pi^{\mp}/D^{* \pm}}=3.40\pm0.42\stat ^{+0.78}_{-0.63}\syst \%$
and
$R_{D_2^{* 0}\to D^{* \pm}\pi^{\mp}/D^{* \pm}}= 1.37\pm0.40\stat^{+0.96}_{-0.33}\syst \%$.
Extrapolating to the full kinematic phase space by a              MC  simulation           
         and using the          partial width ratio,
  $\Gamma(\dtstz \to \dplus \pi^-) /
 \Gamma(\dtstz \to \dstarplus \pi^-)=2.3\pm0.6$~\cite{PDG},
                              the rate of c quarks hadronising
as $\dstarplus$ mesons,
 $f(c \to \dstarplus)=0.235\pm0.007\pm0.007$~\cite{lgcomp},
                                and isospin conservation, 
                    the rates of c quarks hadronising as
$D^0_1$ and $D^{*0}_2$ mesons are found to be:                                        
$f(c\to~D^0_1)=                             
   1.46\pm0.18\stat^{+0.33}_{-0.27}\syst \pm0.06 \ext \%$ and
$f(c\to~D^{*0}_2)=                             
   2.00\pm0.58\stat^{+1.40}_{-0.48}\syst \pm0.41\ext \%$.       The third   
errors arise from   uncertainties in                                                   
 $f(c \to \dstarplus)$ and                                                                
                                the $D^{*0}_2\to D^{*+}\pi^-$ branching ratio.
The results are consistent with                           
                  the           $e^+ e^-$ rates measured  
by     CLEO~\cite{CLEO}:
$f(c\to D^0_1)=                             
   1.8\pm0.3\%$ and
$f(c\to D^{*0}_2)=                             
   1.9\pm0.3\%$.                        
 
 \vspace*{-0.4cm}                                                   
 
\subsection{Search for radially excited   
$D^{*'\pm}$}                                                             
 
Radially excited charm mesons with mass around
                      $2.6\GeV$ are predicted~\cite{radial}
to decay into  $D\pi\pi$ or $D^*\pi\pi$.                                  
A narrow resonance in the  
 $D^{*\pm}\pi^+\pi^-$ final state at $2637\MeV$, interpreted as the radially excited 
$D^{*'\pm}$, was reported by DELPHI~\cite{DELPHI}. No evidence for this
state has been found by OPAL and CLEO~\cite{OPALCLEO}.
 
$D^{*'\pm}$  candidates were reconstructed
                       by ZEUS~\cite{pwave} from their decays to
         $D^{*\pm}\pi^+_4\pi^-_5$.    
No narrow resonance is seen in
                                  the extended mass difference
$M(K\pi\pi_S\pi_4\pi_5)-M(K\pi\pi_S)+M(\ds)$.  An upper limit of
 $R_{\dstprplus \to \dstarplus \pi^+\pi^-/\dstarplus}<2.3\%~~(95\%~~C.L.)$ is obtained
in the measured kinematic region for
   the fraction of $\dspm$        originating from    
$D^{*'\pm}$ decays          within a signal window    $2.59 <
M(D^{*'\pm}) < 2.67\GeV$, which covers theoretical predictions~\cite{radial} and the
DELPHI measurement~\cite{DELPHI}.
Extrapolating by a MC simulation to the full kinematic phase space and using the known $f(c\to D^*)$ value,
          a   
$D^{*'\pm}$ production limit of                                                             
 $f(c \to \dstprplus) \cdot B_{\dstprplus \to \dstarplus \pi^+ \pi^-}
                                          < 0.7 \%~(95\%~C.L.)$ is obtained.
A similar limit of  $0.9\%$ has been reported    by OPAL~\cite{OPAL}.
 
 \vspace*{-0.4cm}                                                   
 
\subsection{Production of the charm-strange meson $D_{s1}^{\pm}(2536)$}
 
$D_{s1}^{\pm}$ mesons were reconstructed by ZEUS~\cite{ds1}
 via the             $\dspm K^0_S$ decay mode                
                                   with $K^0_S\to\pi^+\pi^-$.
$K^0_S$ candidates were identified by using pairs of oppositely charged tracks with
$\pt > 0.2\GeV$.
                                  A      clean                                         
                                               $K^0_S\to \pi_3\pi_4$                  
                          signal was      extracted after
applying standard         $V^0$-finding cuts.  $K^0_S$ candidates with
$0.480 < M(\pi_3\pi_4) < 0.515\GeV$ were kept for the $D_{s1}^{\pm}$  reconstruction.
                                          Fig.~5(d)   shows the          
effective $M(\dspm K^0_S)$ distribution in terms of
$\Delta M^{ext}+M(D^{*+})_{PDG}+M(K^0)_{PDG}$              (solid dots), where
$\Delta M^{ext}=
 M(K\pi\pi_S\pi_3\pi_4)-M(K\pi\pi_S)-M(\pi_3\pi_4)$  and                     
$M(D^{*+})_{PDG}$ ($M(K^0)_{PDG}$) is the nominal
                       $\dspm$ ($K^0$) mass~\cite{PDG}.
A clear signal is seen at the $M(D^{\pm}_{s1})$ value.
                        The solid curve is an unbinned likelihood
fit to a Gaussian resonance plus background of the form $A(\Delta M^{ext})^B$. The fit
yielded $62.3\pm~9.3~D^{\pm}_{s1}$ mesons. The mass value was found to be
$M(D^{\pm}_{s1})=2534.2\pm 0.6\pm 0.5\MeV$, in rough agreement with the PDG value~\cite{PDG}.    
The last error is due to the uncertainty in    
$M(D^{*+})_{PDG}$.                                                                       
 
The angular distribution of the 
$D_{s1}$ signal was studied via the                                                       
 helicity angle, $\alpha$,                      between the $K^0_S$ and $\pi_S$
momenta in the $\dspm$ rest frame. The $dN/d\cos\alpha$ 
            distribution                  was fitted to
($1+R\cos^2\alpha$). An  unbinned likelihood fit yielded 
$R = -0.53 \pm 0.32 \stat^{+0.05}_{-0.14}\syst $, consistent with the CLEO
value~\cite{CLEO2}
 $R = -0.23^{+0.40}_{-0.32}$. Both measurements are consistent with $R=0$, i.e.               
$J^P=1^+$ for the 
$D_{s1}$ meson. However, our result is not inconsistent with $R=-1$, 
        i.e. $J^P=1^-$ or $2^+$~\cite{godfrey}.
 
The fraction of $\dspm$ mesons originating from 
$D_{s1}^{\pm}$        in the measured kinematic region is                                         
$R_{D_{s1}^{\pm}\to D^{* \pm}K^0/D^{* \pm}}=1.77\pm0.26\stat^{+0.11}_{-0.09}\syst \%$.
              The rate of c quarks hadronising as
$D_{s1}^+$ mesons, after a MC extrapolation to the             
                    full kinematic phase space, is             
       $f(c\to D^+ _{s1})
  =1.24\pm 0.18\stat^{+0.08}_{-0.06}\syst \pm 0.14\br \%$.
The third error is due to uncertainties in $f(c\to D^{*+})$ and the
$D_{s1}^+\to                                                                                  
  D^{*+}K^0$ branching ratio. The       rate agrees      with the OPAL value~\cite{OPAL2}
   $1.6\pm 0.4\pm 0.3\%$.
This rate   
       is about twice that expected,    assuming
  $f(c\to D^0 _1)\approx 2\%$~\cite{pwave} and $\gamma_s\approx 0.3$~\cite{D_s}, where
$\gamma_s$ is the strangeness suppression factor in charm production.

 \newpage
 \vspace*{-1.8cm}                                                   
\section{Open Beauty Production}

\begin{figure}
 
 \vspace*{-2.2cm}                                                   
 \hspace*{-0.2cm}                                                   
 
  \resizebox{11.5pc}{!}{\includegraphics{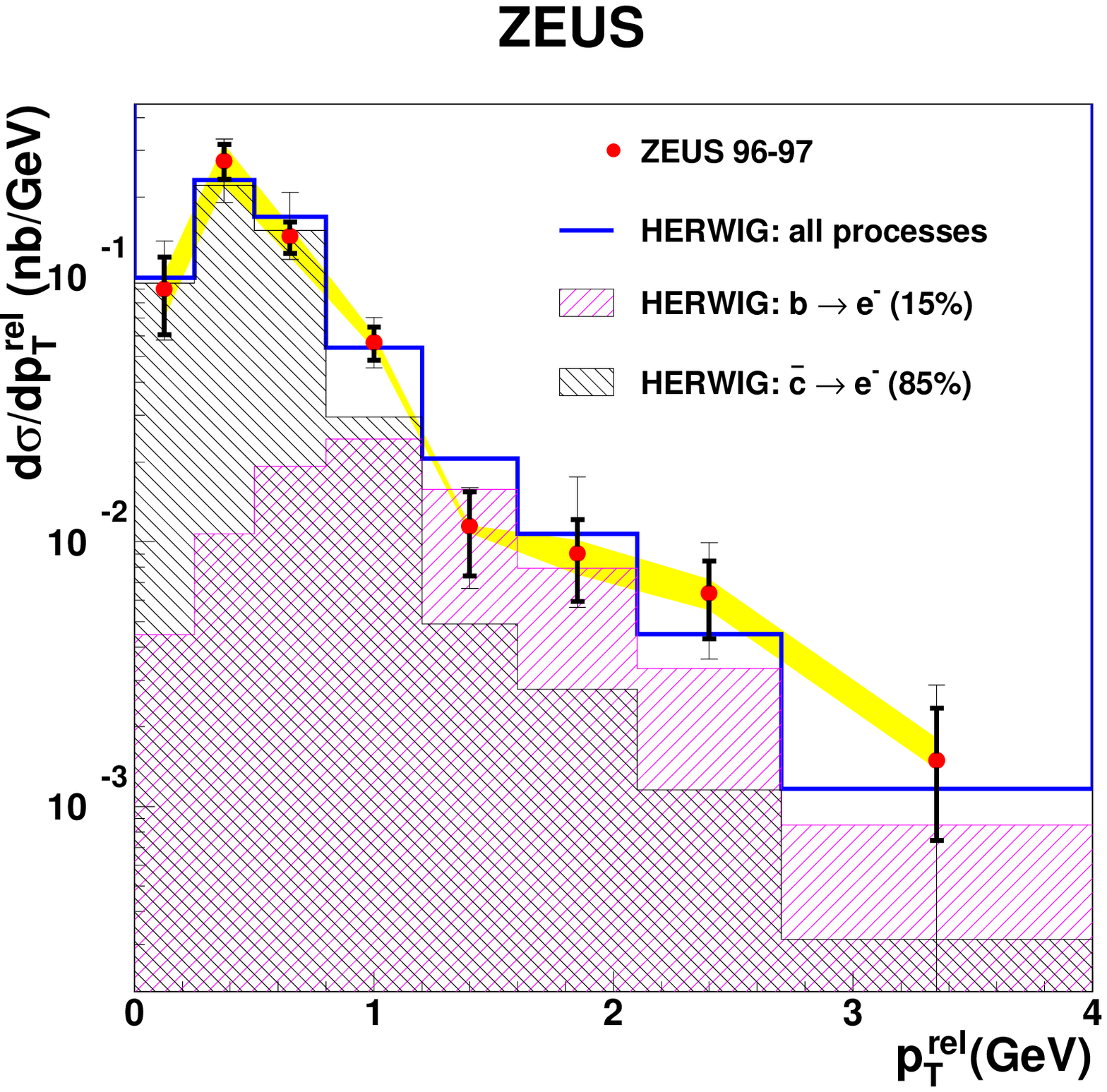}}
\vspace*{2.0cm}\hspace*{ 0.1cm}\resizebox{11.5pc}{!}{\includegraphics{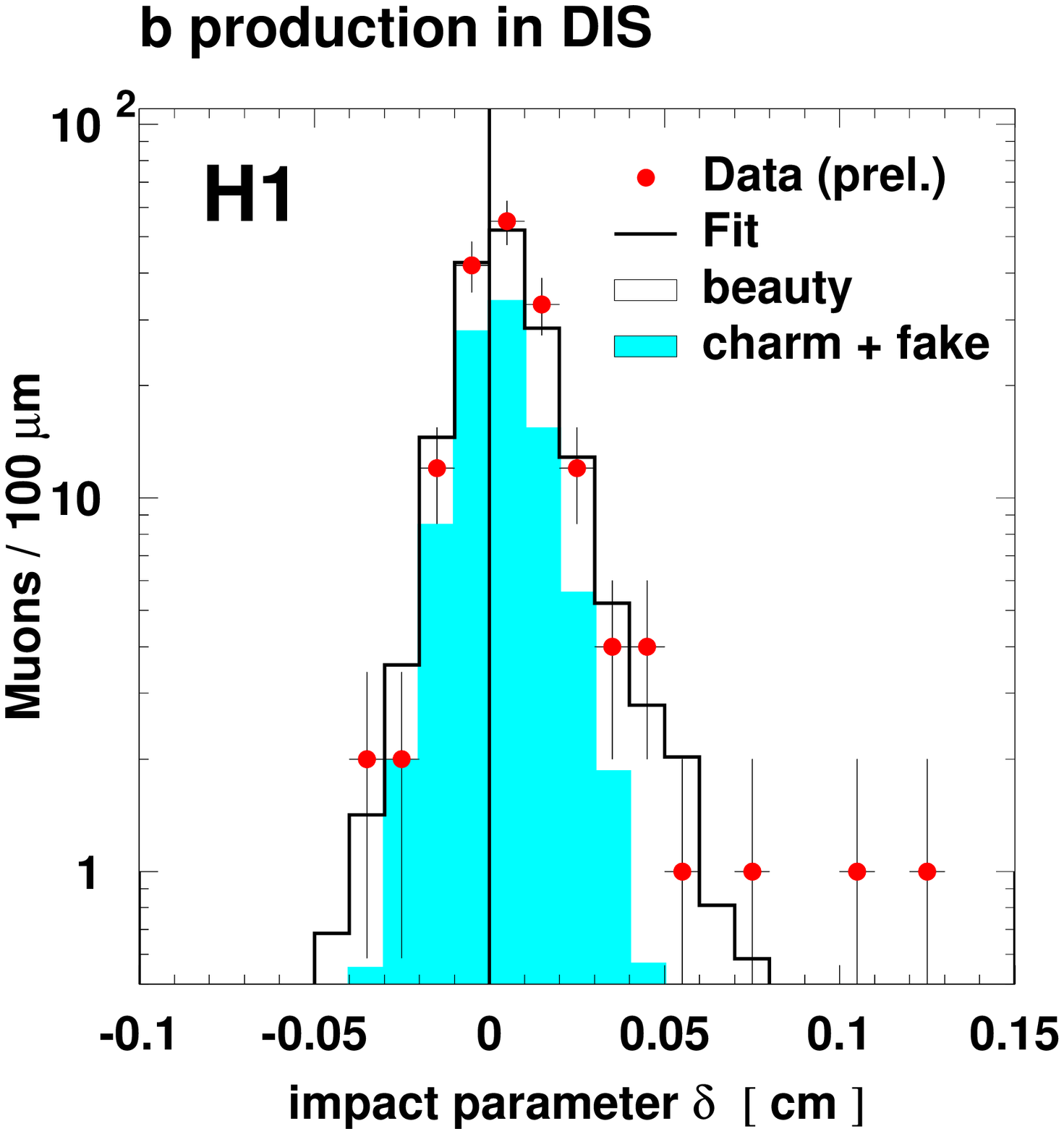}}  
\hspace*{-0.9cm}
 
\vspace*{2.1cm}\hspace*{ 0.8cm}
   \resizebox{11.5pc}{!}{\includegraphics{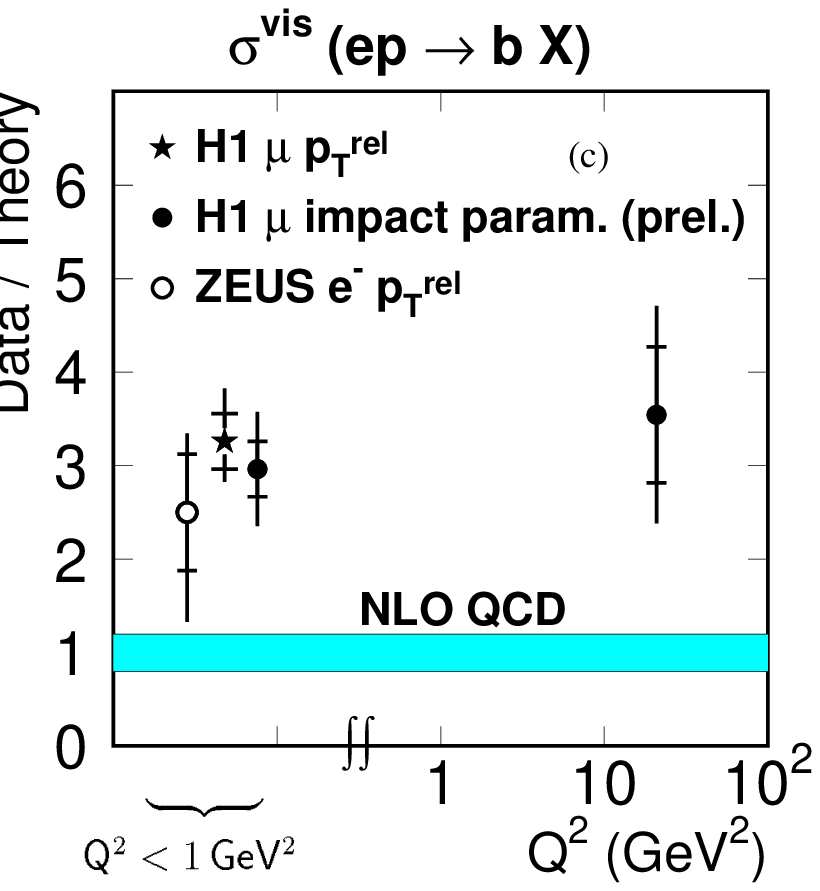}}
\caption{(a)                                             
Differential cross section in $p_T^{rel}$ for the production of 
electrons in dijet PHP events compared to HERWIG MC shape prediction
(solid line). The MC beauty (charm) component is given by the                           
 forward- (backward-) diagonally hatched histogram.
 (b) Muon impact parameter, $\delta$, distribution for DIS events, with a likelihood fit
decomposition into beauty (white region) and charm + fake (shaded region).
 (c) Ratio of measured $b$ production cross sections over theoretical expectation,
as a function of $Q^2$. The shaded band represents the theoretical uncertainty.
        }
\end{figure}
 
 \vspace*{- 8.7cm}\hspace*{ 1.3cm}{\tiny  (a)}
 \vspace*{- 9.9cm}\hspace*{ 4.8cm}{\tiny  (b)}
 \vspace*{-10.0cm}\hspace*{ 9.8cm}{        (c)}
 \vspace*{+28.6cm}\hspace*{-15.9cm}      
 
 \vspace*{-1.0cm}                       
 
The production rates for beauty at HERA are about two orders of magnitude
smaller than for charm. Theoretical uncertainties are expected to be smaller due to
the high $b$ quark mass. Enriched $b$  samples have been obtained~\cite{bH1,bZEUS,bH1del,bnlo}
by studying electrons or muons from semileptonic (SL) $b$ decays. Due to the heavy
$b$ quark, the high tail of the lepton $p_T$ with respect to the axis of the closest jet,
$p_T^{rel}$, provides a $b$ signature. H1 has also exploited, using their micro-vertex      
detector,     the long lifetimes of $b$ hadrons to            
  extract   $b$ production cross sections by measuring the impact parameter, $\delta$,
which is the distance of closest approach of SL muons to the primary vertex 
                    in the plane transverse to
the beam axis.                                                                            
 
The ZEUS differential cross section, $d\sigma /dp_T^{rel}$, for
a PHP dijet event sample of ${\cal L} = 38.5\ipb$                                        
                                     with identified   electrons    
                                                  is compared in Fig.~6(a)~\cite{bZEUS}
                                            with a MC $c$ and $b$ production simulation. The
shape is fitted by varying the relative contributions of $c$ and $b$. A $b$    fraction
of $\approx 15\%$ is obtained, consistent with the MC expectation.
 
For a PHP dijet sample with an identified muon, H1 has shown~\cite{bH1del} that the $p_T^{rel}$ and
    $\delta$ methods provide independent and consistent $b$ production results.
The H1 analysis of
 a smaller DIS sample from ${\cal L} = 10.5\ipb$~\cite{bnlo}  uses      the combination of the
two observables. A likelihood fit of $b\bar b$, $c\bar c$ and fake muon spectra to a
two-dimensional distribution in $\delta$ and 
$p_T^{rel}$ adjusts the relative weights of all three components in the data.             
It yields a $b\bar b$ fraction of $(43\pm 8)\%$. The $\delta$ projection of this distribution
is shown in Fig.~6(b) together with the fit decomposition.
 
The ratio of the measured visible $b$ cross sections over
      theoretical expectations~\cite{hvqdis,frix}
                            is shown in Fig.~6(c) as a function of $Q^2$~\cite{bnlo}. The ratio is roughly
constant with $Q^2$ and the discrepancy between data and NLO calculations is quite significant.
The DIS case is theoretically cleaner, since at high $Q^2$ the resolved contribution is
expected to be suppressed.
 
\doingARLO[\bibliographystyle{aipproc}]
          {\ifthenelse{\equal{\AIPcitestyleselect}{num}}
             {\bibliographystyle{arlonum}}
             {\bibliographystyle{arlobib}}
          }

\newpage
 
 \begin{thebibliography}{99}
 
\bibitem{Frixione}
S.~Frixione et al., Nucl. Phys. B412~(1994)~225;\\
M.~Mangano et al., Nucl. Phys. B373~(1992)~295.  
 
\bibitem{Ellis}     R.K. Ellis and P. Nason, Nucl. Phys. B 312 (1989) 551;\\
                    P. Nason, S. Dawson and R.K. Ellis, Nucl. Phys. B 303 (1988) 607;\\
                    J. Smith and W.L. van Neerven, Nucl. Phys. B 374 (1992) 36.
 
\bibitem{kniehl}  B.A. Kniehl et al., Z. Phys. C76 (1997) 689;\\  
  J.Binnewies et al., Z. Phys. C76 (1997) 677;
 Phys. Rev. D58 (1998) 014014.  
 
\bibitem{cacciari} M. Cacciari et al., Z. Phys. C69 (1996) 459; 
                                     Phys. Rev. D55 (1997) 2736;\\ 
                    M. Cacciari and M. Greco, Phys. Rev. D55 (1997) 7134.  
 
\bibitem{ds1}                                                   
ZEUS Collaboration, paper 497 submitted to    
              EPS Conf. on HEP2001,
Budapest, Hungary.                                   
 
\bibitem{pv}                                                   
ZEUS Collaboration, paper 501 submitted to    
              EPS Conf. on HEP2001,
Budapest, Hungary.                                   
 
\bibitem{pvlep} OPAL Collaboration, K.~Ackerstaff et al.,      
Eur. Phys. J. C 5 (1998) 1; \\
              ALEPH  Collaboration, R.~Barate et al.,      
Eur. Phys. J. C 16 (2000) 597. 
 
\bibitem{D_s}                                                  
ZEUS Collaboration, J.~Breitweg et al.,                                  
 Phys. Lett. B481 (2000) 213.
 
\bibitem{lgcomp} L. Gladilin,                 
hep-ex/9912064. 
 
\bibitem{hvqdis} B.W. Harris and J. Smith,                      
                                     Phys. Rev. D57 (1998) 2806;\\ 
E. Laenen et al.,                    Nucl. Phys. B392 (1995) 162.  
 
\bibitem{osa855}      
   ZEUS Collaboration, paper 449 submitted to XXX      
ICHEP2000, Osaka, Japan.     
 
\bibitem{bu493}                                                 
ZEUS Collaboration, paper 493 submitted to    
              EPS Conf. on HEP2001,
Budapest, Hungary.                                   
 
\bibitem{H1dis}                                                 
H1   Collaboration, C.~Adloff et al., hep-ex/0108039, submitted to     
 Phys. Lett. B (2001).
 
 
 \bibitem{cascade}
 H.~Jung and G.P.~Salam,                                               
Eur. Phys. J. C19 (2001) 351;   
 H.~Jung,     hep-ph-0109146.      
 
 \bibitem{CCFM}
M.~Ciafaloni,  Nucl. Phys. B296 (1988) 49;\\     
S.~Catani et al.,  Phys. Lett. B234 (1990) 339;    Nucl. Phys. B 336 (1990) 18;\\     
G.~Marchesini,  Nucl. Phys. B445 (1995) 49.       
 
\bibitem{dstarrf}ZEUS Collaboration, Breitweg J. et al.,      Eur. Phys. J. 
       C6 (1999) 67.
 
\bibitem{frix}S.~Frixione et al.,      Nucl. Phys. 
       B454 (1995) 3 ;   Phys. Lett.
       B348       (1995) 633.
 
\bibitem{BKL} A.V.~Berezhnoy, V.V.~Kiselev, and A.K.~Likhoded,
  hep-ph/9901333, hep-ph/9905555, Yad.\ Fiz.\ [Phys.\ At.\ Nucl.] in
  print (2000).
 
\bibitem{ben}
ZEUS Collaboration, paper 495 submitted to    
              EPS Conf. on HEP2001,
Budapest, Hungary.                                   
 
\bibitem{aroma}
G. Ingelman et al., Comp. Phys. Comm. 101 (1997) 135.
 
 \bibitem{DGLAP}
V.~Gribov and L.~Lipatov, Sov. J. Nucl. Phys. 15 (1972) 438;           
                                        ibid.  15 (1972) 675;\\     
 L.~Lipatov, Sov. J. Nucl. Phys. 20 (1975) 94;\\     
G.~Altarelli and G.~Parisi, Nucl. Phys. B126 (1977) 298;\\     
Y.~Dokshitser, Sov. Phys. JETP 46 (1977) 641.     
 
\bibitem{cost}
ZEUS Collaboration, paper 499 submitted to    
              EPS Conf. on HEP2001,
Budapest, Hungary.                                   
 
\bibitem{PDG} Particle Data Group,                             
         D. E. Groom et al.,      Eur. Phys. J  C15 (2000) 1.  
 
\bibitem{HQET} N. Isgur and M.B. Wise, Phys. Lett. B232 (1989) 113; \\
         M. Neubert,      Phys. Rep. A245 (1994) 259.
 
\bibitem{pwave}                                                              
   ZEUS Collaboration, paper 448 submitted to XXX      
ICHEP2000, Osaka, Japan.     
 
\bibitem{CLEO}CLEO Collaboration, Avery P. et al.,      Phys. Lett. 
\      B331 (1994) 236.
\bibitem{radial}S. Godfrey and N. Isgur, Phys. Rev. D32 (1985) 189; \\
                D. Ebert et al.,         Phys. Rev. D57 (1998) 5663.   
 
\bibitem{DELPHI} DELPHI Collaboration, P. Abreu et al.,                  
                            Phys. Lett. B426 (1998) 231.    
 
\bibitem{OPALCLEO} OPAL Collaboration, XXIX                    
        ICHEP1998, Vancouver, Canada; \\
                   CLEO Collaboration, hep-ex/9901008.                   
 
\bibitem{OPAL} OPAL Collaboration, hep-ex/0101045, submitted to              
                                Eur. Phys. J.~C (April 2001).
 
\bibitem{CLEO2} CLEO Collaboration, J.P. Alexander et al.,    
                            Phys. Lett. B303 (1993) 303.    
 
\bibitem{godfrey}    
     S. Godfrey and R. Kokoski,            
                                         Phys. Rev. D43 (1991) 1130.   
 
\bibitem{OPAL2} OPAL Collaboration, K. Ackerstaff et al.,                                     
                      Z. Phys. C76 (1997) 425.
 
\bibitem{bH1}
H1 Collaboration, 
C.~Adloff et al., Phys. Lett. B467~(1999)~156.                                      
 
\bibitem{bZEUS}
 ZEUS Collaboration, J. Breitweg et al., 
                               Euro. Phys. J. C18~(2001)~625.
 
\bibitem{bH1del}                                                             
   H1   Collaboration, abstract 982 submitted to XXX      
ICHEP2000, Osaka, Japan.     
 
 
\bibitem{bnlo}
H1   Collaboration, paper 807 submitted to    
              EPS Conf. on HEP2001,
Budapest, Hungary.                                   
 
 
\end{thebibliography}

\end{document}